\documentstyle[prb,aps,multicol,epsf]{revtex}
\input psfig
\def \be{\begin{equation}}
\def \ee{\end{equation}}

\def \o{\omega}

\def \ve{\varepsilon}

\def \s1{\sigma_1}
\begin{document}

\bibliographystyle{simpl1}

\title{Loss of $\pi$-Junction Behaviour in an Interacting Impurity 
Josephson Junction}

\author{Aashish A. Clerk and Vinay Ambegaokar}
\address{Laboratory of Atomic and Solid State Physics,
Cornell University, Ithaca NY 14853, USA}
\maketitle
\centerline{December 20, 1999}
\begin{abstract}

Using a generalization of the non-crossing approximation which
incorporates Andreev reflection, we study the properties of an infinite-U
Anderson impurity coupled to two superconducting leads.  In the regime where
$\Delta$ and $T_K$ are comparable, we find that the position 
of the sub-gap resonance in
the impurity spectral function develops a strong anomalous phase dependence--
its energy is a minimum when the phase difference between the 
superconductors is equal to $\pi$.
Calculating the Josephson current through the impurity, we find that
$\pi$-junction behaviour is lost as the position of the 
bound-state moves above the Fermi energy.  

\end{abstract}
\begin{multicols}{2}

What is the Josephson current through an interacting Anderson 
impurity?  As this 
phenomenon involves the coherent transport of pairs of electrons through
the impurity, one would expect that a strong on-site repulsion would greatly
diminish the effect.  Though this intuitive expectation is partially
correct, there is a variety of non-intuitive behaviour 
also associated with this system.  

If the on-site repulsion is large and the impurity is
singly occupied, a lowest-order perturbative calculation
in the impurity-superconductor coupling reveals that the sign of the 
Josephson coupling can become negative, meaning that the corresponding 
ground state of the system has a phase difference of 
$\delta=\pi$ between the
two superconductors \cite{MATVEEV} \cite{KIVELSON}.
An appealing explanation for this behaviour was provided by 
Spivak and Kivelson  \cite{KIVELSON}, who 
showed that in the limit of single occupancy, it is impossible to
transport a pair across the impurity while preserving its spin ordering,
leading to an extra factor of $(-1)$.
More generally, $\pi$-junction behaviour is expected whenever spin-flip
tunneling processes dominate \cite{KULIK} \cite{SHIBA}.  
Recent work using a Hartree-Fock type procedure indicates this behaviour
can survive beyond lowest-order perturbation theory \cite{ROZHKOV}.

Alternatively, in the limit of large impurity-superconductor coupling, 
the physics of the Kondo effect will also become significant.  
A resulting enhancement of the Josephson current through the impurity 
has been 
predicted in the regime where the superconducting gap $\Delta$ is much
smaller than the Kondo temperature $T_K$ \cite{MATVEEV}.  Significantly,
no $\pi$-junction behaviour is expected here, as the spin of the impurity
is completely screened by the Kondo effect.  

Given these prior results, it is natural to ask 
how large the ratio $T_K/\Delta$ must be to see the destruction of
$\pi$-junction behaviour, and what the properties of the junction
are in this cross-over region; these questions are the motivation for the 
current paper.
We examine the regime where $\Delta$ is comparable in magnitude
to $T_K$, a regime in which the effects of both Kondo physics and 
superconductivity must be treated on an equal footing.  
Note this parameter
range is also of interest as it is more appropriate to mesoscopic
quantum dot systems, which would be one possible experimental realization of
the model.   
We find, somewhat surprisingly,  that the properties of our 
system can be understood in terms of another well-studied feature
of magnetic impurities coupled to superconductors-- the sub-gap bound state.
Typically, the position of this state is a function of the ratio
$T_K/\Delta$ \cite{MULLERH,MATSUURA,JARRELL,CHUNG}.  
We find now that this resonance also develops a pronounced
dependence on the phase difference-- as $\delta$ increases from zero, the 
energy of the bound state {\it decreases}.  This phase dependence
is anomalous in the sense that it indicates an energetic preference away from $\delta = 0$.
Interpreting this sub-gap
state as a current carrying Andreev bound state \cite{BEENAKKER}, we are
lead to the conclusion that the system is a $\pi$-junction if the 
sub-gap bound state is located below the Fermi energy $E_F$.  This relation is
confirmed by making an explicit calculation of the Josephson current through
the impurity.      

{\it Formalism.}--The model we study consists of an 
infinite-$U$ Anderson impurity coupled to two superconducting leads,
each having a different phase.  Using a slave-boson representation, we have:  

\begin{eqnarray}
H = H_0+\varepsilon _d \sum _{\sigma } f^{\dag}_{\sigma }
f^{\phantom{\dag}}_{\sigma } + W\sum _{\alpha ,k,\sigma} 
(c^{\dag}_{\alpha ,k\sigma}b^{\dag } f^{\phantom{\dag}}_{\sigma} + h.c.),
\label{HDOT}
\end{eqnarray}
\begin{eqnarray}
H_0 = \sum _{\alpha,k,\sigma} \varepsilon _k 
c_{\alpha,k\sigma}^{\dag}c^{\phantom{\dag}}_{\alpha,k\sigma} + 
\sum _{\alpha, k} 
(\Delta_{\alpha} c^{\dag}_{\alpha,k\uparrow} c^{\dag}_{\alpha,-k\downarrow} + h.c. ).
\label{HO}
\end{eqnarray}
{\noindent}The $c^{\dag}_{\alpha ,k\sigma}$ operators here
create band electrons, with 
$\sigma$ denoting spin and $\alpha =L,R$ labelling the left and right
superconducting leads.  
$\Delta_{\alpha}=|\Delta| \exp(i\phi_{\alpha})$ represents the pair 
potential in lead $\alpha$, with the phase difference $\delta$ being
defined as $\phi_R - \phi_L$.  The
Anderson impurity has bare energy $\varepsilon _d$, and is represented
in the usual manner using auxiliary fermion ($f$) and boson ($b$) operators;
the $U = \infty$ constraint of single occupancy takes the form 
$\sum _{\sigma } f^{\dag}_{\sigma }f^{\phantom{\dag}}_{\sigma } + b^{\dag}b
= 1$. 

{\it The NS-NCA}.--We calculate the impurity spectral function 
(also called the impurity density of states) for our system using the
NS-NCA \cite{AASH}, an 
extension of the self-consistent non-crossing approximation (NCA) 
\cite{BICKERS} to superconducting systems.  The NCA
amounts to an infinite order re-summation of perturbation theory,  and 
has been shown to be quantitatively reliable 
down to temperatures below $T_K$ \cite{BICKERS}\cite{NCACAVEAT}.  
The modification we employ self-consistently
 includes multiple-Andreev reflection 
processes in a manner which is formally exact to order $1/N$, 
where $N=2$ is the spin
degeneracy of the impurity.
As the success of the normal NCA is attributed to the fact
that it too is exact to order $1/N$ (in the absence of superconductivity), 
the NS-NCA 
employed here can be viewed as a natural extension to systems with
superconductivity.
The graphs determining the $f$-fermion and $b$ slave boson propagators
are given in Fig. 1.
\begin{figure}[t]
\centerline{\psfig{figure=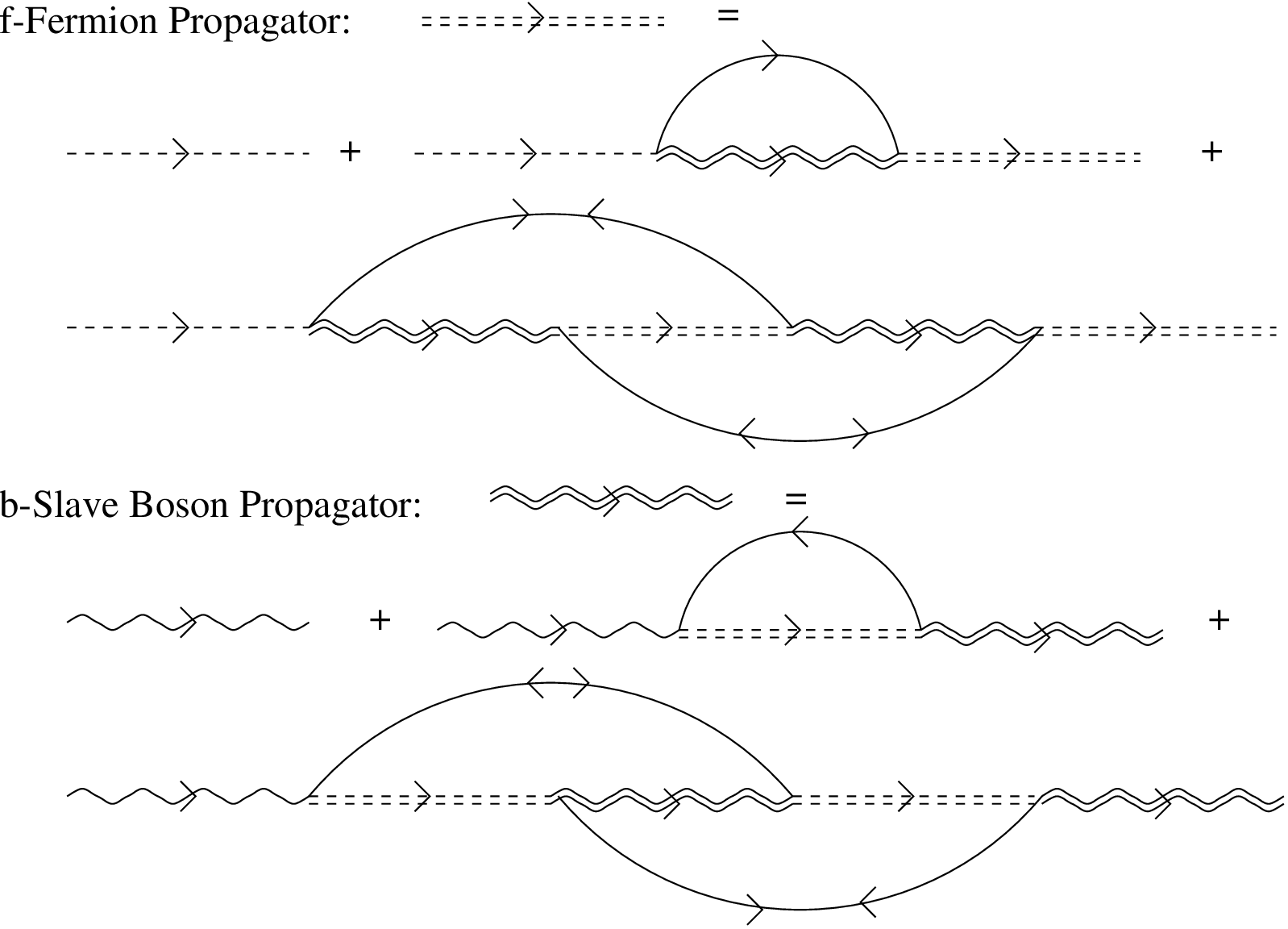,width=6.1cm}}
\narrowtext
\caption{
Diagrammatic representation of the NS-NCA.  Dashed lines are f-fermions,
wavy lines are slave bosons, solid lines are lead electrons. Double lines
indicate a fully dressed propagator.  Anomalous propagators indicate 
Andreev reflection.
}
\label{nsnca}
\end{figure}
Note that the anomalous graphs appearing in Fig. (\ref{nsnca}) always
involve {\it two} Andreev reflections.  As these Andreev reflection events
pick up the phase of the superconductor in which they occur, and each
Andreev reflection is free to occur in either of the two superconductors,
the phase difference $\delta$ naturally enters the impurity 
self energies through an interference term.  A previous study using the NCA
to study the regime $\Delta >> T_K$ 
\cite{SONE} neglected these diagrams, and thus 
omitted the only phase-dependent contribution to the impurity Green function 
which survives in the limit of a strong on-site repulsion.

The Dyson equations pictured diagrammatically in Fig. 1 lead to a set of
coupled integral equations for the $f$-fermion and slave boson propagators. 
Letting $F(\omega ) =(\omega -\varepsilon _d - 
\Sigma (\omega ))^{-1}$ and $B(\omega ) 
=(\omega - \Pi(\omega ))^{-1}$ represent the $f$-fermion and slave boson 
retarded propagators respectively, the equations read:

\begin{eqnarray}
\Sigma(\omega) = \frac{\Gamma}{\pi} \int d\varepsilon
\biggl( \rho (\epsilon) B(\omega - \varepsilon) f(-\varepsilon) -
\label{NSF}
\\
\frac{\Gamma}{\pi} \int d\varepsilon' 
\tau(\varepsilon) \tau(\varepsilon') 
B(\omega + \varepsilon) F(w + \varepsilon + \varepsilon')
B(\omega + \varepsilon') \biggr)
\nonumber
\end{eqnarray}

\begin{eqnarray}
\Pi(\omega) = \frac{2\Gamma}{\pi} \int d\varepsilon
\biggl( \rho (\epsilon) F(\omega + \varepsilon) f(\varepsilon) +
\label{NSB}
\\
\frac{\Gamma}{\pi} \int d\varepsilon' 
\tau(\varepsilon) \tau(\varepsilon') 
F(\omega + \varepsilon) B(w + \varepsilon + \varepsilon')
F(\omega + \varepsilon') \biggr).
\nonumber
\end{eqnarray}

In these equations, 
$\rho(\varepsilon)$ is the electronic density of 
states, $\Gamma=\pi W^2 \rho(0)$ the 
bare tunneling rate, $f$ the Fermi distribution function,  
and $\tau(\varepsilon)$ is an
effective electron-hole coherence parameter defined by:
\begin{eqnarray}
\tau(\omega) = \sum _{k,\alpha} u^{*}_{k,\alpha} 
v^{\phantom{*}}_{k,\alpha}
\delta (|\omega| - \varepsilon _{n}) f(\omega)
\label{alph}
\end{eqnarray}
where $u_{k,\alpha}$ and $v_{k,\alpha}$ are the usual BCS coherence factors. 

{\it Results.}-- We use a Gaussian density states
for the band electrons having half-width $D$.  For the results shown
here, we choose model parameters $\varepsilon _d = -0.67D$ and $\Gamma=0.15D$,
which yields an approximate $T_K = 0.0005 D$ \cite{TKDEFN}.  
We have solved numerically via
iteration the NS-NCA equations in equilibrium for various temperatures,
gap sizes and phase differences.  Within the NCA, the 
impurity spectral function $A_d(\o)$ can be directly related to the $f$-fermion
and slave boson spectral functions:

\begin{eqnarray}
A_{d\sigma}(\omega) = \int {d\varepsilon} [{\rm e}^{-\beta\varepsilon} + 
{\rm e}^{-\beta(\varepsilon - \omega)}] 
A_{f\sigma}(\varepsilon)A_{b}(\varepsilon - \omega)
\label{AD}
\end{eqnarray}
where the auxiliary particle spectral functions are 
defined by $A_{f} = - \frac{1}{\pi} {\rm Im} \, F$, 
$A_{b} = - \frac{1}{\pi} {\rm Im} \, B$. 
{\noindent} Note that the equality in equation (\ref{AD}) reflects a neglect
of vertex corrections which is consistent with the large-$N$ nature of the
NCA.  
\begin{figure}[t]
\centerline{\psfig{figure=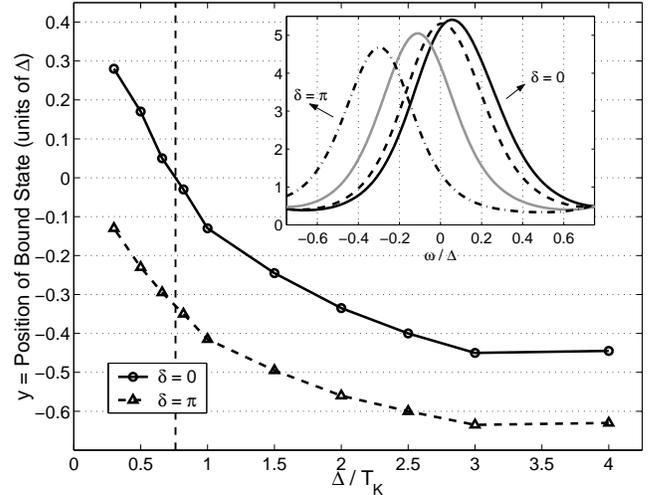,width=8.5cm}}
\narrowtext
\caption{
Position of sub-gap bound state as a function of $\Delta/T_K$ for a
fixed $T/\Delta = 0.5$, $T_K = 0.0005D$. 
The vertical dashed line indicates the approximate value of $\Delta$ for which
the bound state resonance crosses the Fermi energy.  Inset: plots of the 
impurity density of states $A_d(\omega)$ for $\Delta/T_K =0.66$
at various phase differences.  From right to left, we have 
$\delta=0$ (solid curve), $\delta=\pi/4$, $\delta=\pi/2$, $\delta=\pi$ (dash-dot curve). 
}
\label{BNDPOS}
\end{figure}
The solid curve in Fig. \ref{BNDPOS} 
(with points indicated by circles) shows the
position $y$ (in units of $\Delta$) of the sub-gap resonance in the
impurity density of states as a function of $\Delta/T_K$ at $\delta=0$
and at a constant ratio $T/\Delta = 0.5$; this resonance corresponds to the
sub-gap bound state.  As has
been observed in earlier studies, we find that as $\Delta/T_K$ decreases, 
the position of the bound state increases, with the transition 
across the Fermi energy occurring when $\Delta$ is on the order of $T_K$.  Note
that as we are not using a model with particle-hole symmetry, we only have
a single sub-gap state.  

A cogent interpretation of this behaviour 
was provided in \cite{MATSUURA,JARRELL}, where the position of the
sub-gap bound state was suggested to result from the competition between
two possible ground states. For $T_K> \Delta$, the ground state
of the system is the Kondo singlet, and the excited sub-gap
state is a spin doublet. 
Recall that in the Kondo singlet ground state of the infinite-$U$
Anderson model, there is a small probability of finding the impurity empty,
despite the fact that the bare site energy $\ve_d$ is well below $E_F$;  
this is not true for the doublet state, where the impurity is expected
to be singly occupied.
Thus, the fact that the bound state is located {\it above}
 $E_F$ in this limit
(i.e. $y>0$) reflects the fact 
that one must {\it add} a particle at the impurity 
to the singlet ground state to reach 
the doublet state.  The magnitude of $y$ indicates 
the energy splitting between these two states:
\be
y = \frac{E_{excited} - E_{ground}}{\Delta}
  = \frac{ E_{doub} - E_K}{\Delta} > 0
\label{yposeqn}
\ee
{\noindent}where $E_{doub}$ is the energy of the doublet state
and $E_{K}$ is that of the Kondo singlet state.
Similarly, for $\Delta > T_K$, the ground
state is the fully-paired doublet favoured by the superconductivity; the bound
state is located below $E_F$ as one must {\it remove} a particle from the
doublet ground state to reach the excited singlet state.  An
identical expression holds for $y$:
\be
y = \frac{-(E_{excited} - E_{ground})}{\Delta} 
  = \frac{-(E_K - E_{doub})}{\Delta} < 0
\label{ynegeqn}
\ee
{\noindent}The crossing
of $E_F$ by the sub-gap state is thus seen to indicate a substantial
change in the nature of the ground state.

While the behaviour of the bound state at $\delta=0$ has received much 
attention, the effects of having a phase difference has not, to our
knowledge, been previously
studied.  The triangles in Fig. \ref{BNDPOS} indicate 
the bound state positions $y$ for the same values of $T_K / \Delta$ 
as the solid curve, but now with
a phase difference of $\delta=\pi$ between the superconductors.  
We find that for all values tested, $y$ 
{\it decreases} as we increase $\delta$ to $\pi$-- the sub-gap bound state
has an anomalous phase dependence.  This behaviour is seen more explicitly
in the inset of Fig. \ref{BNDPOS}, 
where we plot the sub-gap resonance for various values of $\delta$.     

The phase dependence of the sub-gap state can also be interpreted in
terms of a competition between the Kondo singlet and the fully paired doublet
states.  The Kondo state is again expected to be the ground 
state for $T_K > \Delta$, while the the doublet state will be the ground 
state in the opposite regime.  Significant now however is the fact that
the phase dependence of these two states is quite 
different.  In the doublet state, the impurity can be viewed as a local moment
and we thus expect a negative Josephson coupling and that $E_{doub}$ will
be a {\it minimum} at $\delta=\pi$.  In the Kondo singlet state, 
the impurity spin has been screened and thus we do not expect any anomalous
Josephson coupling-- $E_K$ will be a {\it maximum} at $\delta=\pi$.  With these
associations, it follows from Eqs. (\ref{yposeqn}) and 
(\ref{ynegeqn}) that $y$ {\it is minimized at} $\delta = \pi$, meaning that
the sub-gap state has an anomalous phase dependence.  Note that for
small values of $\Delta/T_K$, it is possible to make the sub-gap state
cross $E_F$ and thus change markedly the nature of the
system ground state by only changing $\delta$. 
     
At this point, it is natural to attempt to make a connection to the Josephson
effect.  Recall that for a non-interacting system, it is possible to discuss
the Josephson current as being at least partially carried by Andreev bound
states existing in the weak link between the two superconductors 
\cite{BEENAKKER}.  These
states have a phase-dependent energy, and thus contribute to the current
through the relation $I = \frac{2e}{\hbar} dF/d\delta$, where $F$ is the free
energy of the junction.  For the simple case where a non-interacting impurity
couples the two superconductors, there are always two Andreev bound
states with energies $\pm |\ve(\delta)|$; as the lower-energy state 
$-|\ve(\delta)|$ has a ``normal'' phase dependence ($dE/d\delta > 0$), the
Josephson current is always positive.

In the present case, matters are quite different due to the strong on-site interaction.  
We have only one apparent Andreev bound state, and its
phase-dependence is anomalous ($dE/d\delta < 0$).  We would thus naively expect that
the Josephson current would both become negative and enhanced in magnitude
as the bound state crosses below $E_F$ and becomes ``occupied''.  Of course, this picture 
is oversimplified,
as there are other possible contributions to the Josephson current-- 
due to interactions, the impurity
contribution to the free energy $F$ is not just a simple function of the impurity density of states.

The conjecture made above can be tested by making an explicit calculation
of the DC Josephson current through the impurity.  Using a technique similar
to that used in \cite{WINGREEN} to calculate the normal current through
an interacting impurity, we arrive at the following exact expression
for the Josephson current \cite{phasedep}:

\begin{eqnarray}
\label{jcurr}
I_{J} =  \frac{4e}{\hbar} \Gamma 
     \sin(\frac{\delta}{2}) & \sum_k & \frac{W^2}{\Gamma} 
\int d\omega f(\omega) \times  \\
\nonumber
 && \frac{-Im}{\pi} \left( g^{R,12}_k(\omega) G^{R,21}_d(\omega,\delta) 
\right).
\label{kkernel}
\end{eqnarray}

{\noindent} Here, $g^{R,12}_k(\omega)$ indicates an anomalous BCS propagator,
and $G^{R,21}_d(\omega,\delta)$ is the anomalous impurity propagator; without
loss of generality, we have chosen $\phi_L + \phi_R = 0$. 

We calculate the anomalous impurity propagator within the NCA using 
the lowest-order contributing graph \cite{BICKERS} \cite{AASH}, without 
making any further approximation.  To calculate an approximate value
of the $0$ temperature Josephson current, we use Eq. \ref{jcurr} with the Fermi
function taken at $T=0$, but with all spectral functions
calculated at $T = 0.5 \Delta$ (the NCA is unreliable at $T=0$
\cite{NCACAVEAT}).  This is a reasonable procedure, as we expect
no qualitative changes in the spectral functions as the
temperature is further lowered;
the sub-gap resonances will only be sharpened.  Retaining a finite $T$
Fermi function would smear the contribution of the sub-gap
resonances, and thus obscure the behaviour we are interested in.  
\begin{figure}[t]
\centerline{\psfig{figure=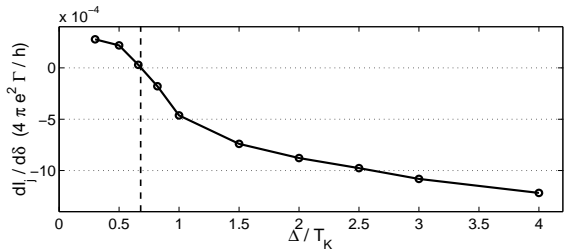,width=7.5cm}}
\narrowtext
\caption{
Approximate $T=0$ phase-derivative of the Josephson current
($d I_J / d \delta$) at $\delta=0$ for various values
of $\Delta/T_K$.  Note the transition from negative to positive
current (indicated by the vertical dashed line) 
coincides with the sub-gap bound state crossing the Fermi energy
(see Fig. \ref{BNDPOS}).
}
\label{JOEPLOT}
\end{figure}
Plotted in Figure \ref{JOEPLOT} 
is the approximate $0$ temperature phase derivative of the 
Josephson current($d I_J / d \delta$) thus obtained as a function of 
$\Delta/T_K$, taken at $\delta = 0$.  Comparison
to Fig. \ref{BNDPOS} indicates that $\pi$-junction behaviour is indeed
lost when the bound state crosses above the Fermi energy, as expected from a
non-interacting picture.  Similar results hold at other values of $\delta$--
($d I_J / d \delta$) changes sign at roughly
the same value of $\Delta/T_K$  at which the bound state crosses 
the Fermi energy.
It would thus appear that the sub-gap bound state
is the entity responsible for carrying the reversed-sign Josephson current
associated with $\pi$-junction behaviour.  

Note that in some respects, our conclusion here resembles what was
found in a study of the Josephson current through a {\it non-interacting}
magnetic insulating layer \cite{TANAKA}; there also, the transition
to $\pi$-junction behaviour is accompanied by an anomalous Andreev
bound state crossing below the Fermi energy.  We stress again that the
present situation is quite different due to the strong on-site repulsion--
there is no {\it a priori} reason guaranteeing that the general 
non-interacting theory relating $I_J$ to Andreev bound states should be 
applicable to the interacting system studied here.  
We find only a single sub-gap state in the impurity density of
states, whereas in the non-interacting case bound states always occur in pairs,
with one member of each having $dE/d\phi < 0$. 

\begin{figure}
\centerline{\psfig{figure=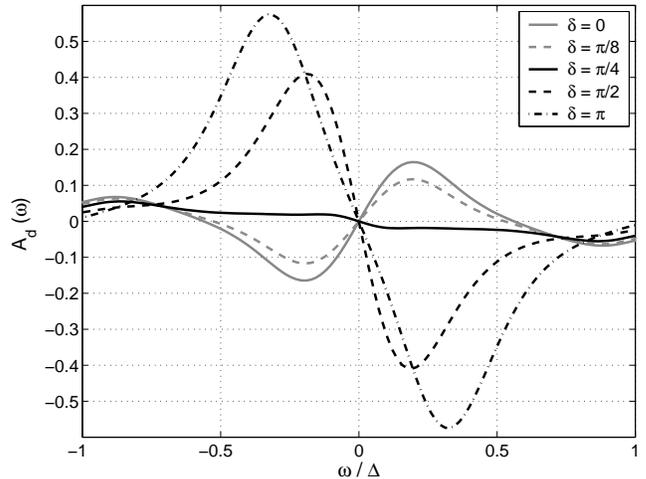,width=8.5cm}}
\narrowtext
\caption{
Anomalous impurity spectral function $Im G^{R,21}_d$
for various values of the
phase difference $\delta$ with fixed $\phi_L + \phi_R = 0$,
$\Delta = 0.66T_K$, $T = 0.5 \Delta$.
Note that as $\delta$ increases, the spectral function changes sign; 
this coincides with the bound state crossing 
below the Fermi energy (see inset of Fig. \ref{BNDPOS}).
}
\label{ANOMSPECT}
\end{figure}

Also of interest in the present system is the behaviour of the 
superconducting phase of the impurity itself.
We find that this phase undergoes a shift by $\pi$ as the bound state passes
through the Fermi energy. When the bound state is above $E_F$, the impurity
simply has the average phase of the two superconducting leads-- at low
energies, the signs of both the real and imaginary parts of the anomalous 
impurity Green function ($G^{R,21}_d$) are the same as those of the average 
anomalous Green functions of the superconductors.  
However, when the bound state moves below $E_F$, we find that
the anomalous impurity green function changes sign.  We interpret this to mean
that the impurity has acquired an additional phase of $\pi$.  This
behaviour is shown Fig. \ref{ANOMSPECT}.  It is clear from
Eq. \ref{kkernel} that this additional factor of $(-1)$ will cause
a sign change in the Josephson current.  A similar
phase shift was found in earlier work involving a single superconductor 
\cite{CHUNG}, though a transition was not observed.

{\it Conclusions.}- Using an extension of the NCA, we have studied
the properties of an infinite-$U$ Anderson impurity coupled to two 
superconductors.  We find that the sub-gap bound state develops an
anomalous phase dependence, and that $\pi$-junction behaviour is lost
when the bound state crosses above $E_F$.

A.C. acknowledges the support of the Olin Foundation.  Work
supported in part by the NSF under grant DMR-9805613.

\end{multicols}


\vskip -0.6cm
\begin{references}
\bibitem{MATVEEV}
L.I. Glazman and K.A. Matveev, JETP Lett. {\bf 49}, 659 (1989).
\bibitem{KIVELSON}
B.I. Spivak and S.A. Kivelson, Phys. Rev. B {\bf 43}, 3740 (1991).
\bibitem{KULIK}
I.O. Kulik, Sov. Phys. JETP {\bf 22}, 841 (1966).
\bibitem{SHIBA}
H. Shiba and T. Soda, Prog. Theor. Phys. {\bf 41}, 25 (1969).
\bibitem{ROZHKOV}
A.V. Rozhkov and D.P. Arovas, Phys. Rev. Lett. {\bf 82}, 2788 (1999).

\bibitem{MULLERH}
J. Zittartz, A. Bringer, and E. Muller-Hartmann, Solid State Commun. 
 {\bf 10}, 513 (1971).
\bibitem{MATSUURA}
T. Matsuura, Prog. Theor. Phys. {\bf 57}, 1823 (1977)
\bibitem{JARRELL}
M. Jarrell, D.S. Sivia, and B. Patton, Phys. Rev. B {\bf 42},
 4804 (1990).
\bibitem{CHUNG}
W. Chung and M. Jarrell, Phys. Rev. Lett. {\bf 77}, 3621 (1996).


\bibitem{BEENAKKER}
C.W.J. Beenakker in {\it Transport Phenomena in Mesoscopic Systems}, edited
 by H. Fukuyama and T. Ando (Springer-Verlag, Berlin, 1992).

\bibitem{AASH}
A.A. Clerk, V. Ambegaokar and S. Hershfield, Phys. Rev. B. {\bf 61},
3555 (2000).
\bibitem{BICKERS}
N. E. Bickers, Rev. Mod. Phys.  {\bf 59}, 845 (1987) and references 
cited within.
\bibitem{NCACAVEAT}
Note that in the $N=2$ case, the NCA does break down at sufficiently low
energy scales ($\simeq T_K^2 / \Gamma <<T_K$); 
we do not push our calculations to the regime
where either $T$ or $\Delta$ are smaller than $0.1 T_K$.

\bibitem{SONE}
S. Ishizaka and J. Sone, Phys Rev. B. {\bf 52}, 8358 (1995).

\bibitem{TKDEFN}
We use the definition $T_K = \sqrt{2 \Gamma} D \exp(\pi \ve_d / (2 \Gamma) ).$

\bibitem{WINGREEN}
Y. Meir and N.S. Wingreen, Phys. Rev. Lett. {\bf 68}, 2512 (1992).
\bibitem{phasedep}
Note that 
$G^{R,21}_d$ will be proportional to $\cos(\delta/2)$, meaning that $I_J$ will
have the expected $\sin(\delta)$ dependence.

\bibitem{TANAKA}
Y. Tanaka and S. Kashiwaya, Physica C, {\bf 274}, 357 (1997).

\end{references}
\end{document}